\documentclass[10pt]{IEEEtran}

\pdfoutput=1

\usepackage{booktabs} 
\usepackage{amsfonts}
\usepackage[dvips]{graphicx}
\usepackage[tight,footnotesize]{subfigure}
\usepackage{listings}
\usepackage{cite}
\usepackage{url}
\usepackage{pifont}
\usepackage{flushend}
\usepackage{algorithm}
\usepackage{algpseudocode}
\usepackage[dvipsnames]{xcolor}
\usepackage{float}
\usepackage{amsmath,xparse}
\usepackage{color, colortbl}

\usepackage{setspace}
\usepackage{booktabs}
\usepackage{footnote}

\definecolor{Gray}{gray}{0.9}
\newcommand{\cparagraph}[1]{\vspace{1.5mm}\noindent\textbf{#1.}}

\makeatletter
\def\therule{\makebox[\algorithmicindent][l]{\hspace*{.5em}\vrule height .75\baselineskip depth .25\baselineskip}}%

\newtoks\therules
\therules={}
\def\appendto#1#2{\expandafter#1\expandafter{\the#1#2}}
\def\gobblefirst#1{
  #1\expandafter\expandafter\expandafter{\expandafter\@gobble\the#1}}%
\def\LState{\State\unskip\the\therules}
\def\pushindent{\appendto\therules\therule}%
\def\popindent{\gobblefirst\therules}%
\def\printindent{\unskip\the\therules}%
\def\printandpush{\printindent\pushindent}%
\def\popandprint{\popindent\printindent}%

\algdef{SE}[WHILE]{While}{EndWhile}[1]
  {\printandpush\algorithmicwhile\ #1\ \algorithmicdo}
  {\popandprint\algorithmicend\ \algorithmicwhile}%
\algdef{SE}[FOR]{For}{EndFor}[1]
  {\printandpush\algorithmicfor\ #1\ \algorithmicdo}
  {\popandprint\algorithmicend\ \algorithmicfor}%
\algdef{S}[FOR]{ForAll}[1]
  {\printindent\algorithmicforall\ #1\ \algorithmicdo}%
\algdef{SE}[LOOP]{Loop}{EndLoop}
  {\printandpush\algorithmicloop}
  {\popandprint\algorithmicend\ \algorithmicloop}%
\algdef{SE}[REPEAT]{Repeat}{Until}
  {\printandpush\algorithmicrepeat}[1]
  {\popandprint\algorithmicuntil\ #1}%
\algdef{SE}[IF]{If}{EndIf}[1]
  {\printandpush\algorithmicif\ #1\ \algorithmicthen}
  {\popandprint\algorithmicend\ \algorithmicif}%
\algdef{C}[IF]{IF}{ElsIf}[1]
  {\popandprint\pushindent\algorithmicelse\ \algorithmicif\ #1\ \algorithmicthen}%
\algdef{Ce}[ELSE]{IF}{Else}{EndIf}
  {\popandprint\pushindent\algorithmicelse}%
\algdef{SE}[PROCEDURE]{Procedure}{EndProcedure}[2]
   {\printandpush\algorithmicprocedure\ \textproc{#1}\ifthenelse{\equal{#2}{}}{}{(#2)}}%
   {\popandprint\algorithmicend\ \algorithmicprocedure}%
\algdef{SE}[FUNCTION]{Function}{EndFunction}[2]
   {\printandpush\algorithmicfunction\ \textproc{#1}\ifthenelse{\equal{#2}{}}{}{(#2)}}%
   {\popandprint\algorithmicend\ \algorithmicfunction}%
\makeatother

\makesavenoteenv{table}
\makesavenoteenv{tabular}

\begin{document}
%
%
\title{Optimizing Sparse Matrix-Vector Multiplication on Emerging Many-Core Architectures \vspace{-2mm}}

\author{\IEEEauthorblockN{Shizhao Chen\IEEEauthorrefmark{2},
Jianbin Fang\IEEEauthorrefmark{2}, Donglin Chen\IEEEauthorrefmark{2}, Chuanfu Xu\IEEEauthorrefmark{1}\IEEEauthorrefmark{2}, Zheng
Wang\IEEEauthorrefmark{3}}

 \IEEEauthorblockA{\IEEEauthorrefmark{1}State Key Laboratory of Aerodynamics, China Aerodynamics Research and
Development Center, China}\\

\IEEEauthorblockA{\IEEEauthorrefmark{2}College of Computer, National University of Defense Technology, China}\\
\IEEEauthorblockA{\IEEEauthorrefmark{3}MetaLab, School of Computing and Communications, Lancaster University, United Kingdom}
\\
Email: \{chenshizhao12, j.fang, chendonglin14, xuchuanfu\}@nudt.edu.cn, z.wang@lancaster.ac.uk

}

\maketitle              

\begin{abstract}
Sparse matrix vector multiplication (SpMV) is one of the most common operations in scientific and high-performance applications, and is
often responsible for the application performance bottleneck. While the sparse matrix representation has a significant impact on the
resulting application performance, choosing the right representation typically relies on expert knowledge and trial and error. This paper
provides the first comprehensive study on the impact of sparse matrix representations on two emerging many-core architectures: the
Intel's Knights Landing (KNL) XeonPhi and the ARM-based FT-2000Plus (FTP). Our large-scale experiments involved over 9,500 distinct
profiling runs performed on 956 sparse datasets and five mainstream SpMV representations. We show that the best sparse matrix
representation depends on the underlying architecture and the program input. To help developers to choose the optimal matrix
representation, we employ machine learning to develop a predictive model. Our model is first trained offline using a set of training
examples. The learned model can be used to predict the best matrix representation for any unseen input for a given architecture. We show
that our model delivers on average 95\% and 91\% of the best available performance on KNL and FTP respectively, and it achieves this with
no runtime profiling overhead.
\end{abstract}

\begin{IEEEkeywords}
Sparse matrix vector multiplication; Performance optimization; Many-Cores; Performance analysis
\end{IEEEkeywords}
\section{Introduction}

Sparse matrix-vector multiplication (SpMV) is commonly seen in scientific and high-performance
applications~\cite{DBLP:conf/ics/SedaghatiMPPS15, DBLP:conf/cf/YangFCWTL17}. It is often responsible for the performance bottleneck and notoriously difficult to
optimize~\cite{DBLP:conf/ics/LiuSCD13, DBLP:conf/ics/0002V15, DBLP:journals/cj/CheXFWW15}. Achieving a good SpMV performance is challenging because its performance is
heavily affected by the density of nonzero entries or their sparsity pattern. As the processor is getting increasingly diverse and complex,
optimizing SpMV becomes harder.




Prior research has shown that the sparse matrix storage format (or representation) can have a significant impact on the resulting
performance, and the optimal representation depends on the underlying architecture as well as the size and the content of the
matrices~\cite{DBLP:conf/sc/BellG09, DBLP:journals/siamsc/KreutzerHWFB14, DBLP:conf/sc/WilliamsOVSYD07, DBLP:journals/fgcs/jing,
DBLP:conf/ipps/ChenFLTCY17}. While there is already an extensive body of study on optimizing SpMV on SMP and multi-core
architectures~\cite{DBLP:conf/ics/LiuSCD13, DBLP:conf/ics/0002V15}, it remains unclear how different sparse matrix representations affect
the SpMV performance on emerging many-core architectures.

This work investigates techniques to optimize SpMV on two emerging many-core architectures: the Intel Knights Landing (KNL) and the Phytium
FT-2000Plus (FTP)~\cite{DBLP:conf/hotchips/Zhang15, ft2000}. Both architectures integrate over 60 processor cores to provide potential high
performance, making them attractive for the next-generation HPC systems. In this work, we conduct a large-scale evaluation involved over
9,500 profiling measurements performed on 956 representative sparse datasets by considering five widely-used sparse matrix representations:
CSR~\cite{DBLP:conf/sc/WilliamsOVSYD07}, CSR5~\cite{DBLP:conf/ics/0002V15}, ELL~\cite{osti_7093021},
SELL~\cite{DBLP:conf/hipeac/MonakovLA10, DBLP:journals/siamsc/KreutzerHWFB14}, and HYB~\cite{DBLP:conf/sc/BellG09}.

We show that while there is significant performance gain for choosing the right sparse matrix representation, mistakes can seriously hurt
the performance. To choose the right matrix presentation, we develop a predictive model based on machine learning techniques. The model
takes in a set of quantifiable properties, or \emph{features}, from the input sparse matrix, and predicts the best representation to use on
a given many-core architecture. Our model is first trained offline using a set of training examples. The trained model can then be used to
choose the optimal representation for any \emph{unseen} sparse matrix. Our experimental results show that our approach is highly effective
in choosing the sparse matrix representation, delivering on average 91\% and 95\% of the best available performance on FTP and KNL,
respectively.

This work makes the following two contributions:

\begin{itemize}
\item It presents an extensible framework to evaluate SpMV performance on KNL and FTP, two emerging many-core architectures for HPC;

\item It is the first comprehensive study on the impact of sparse matrix representations on KNL and FTP;

\item It develops a novel machine learning based approach choose the right representation for a given architecture, delivering
    significantly good performance across many-core architectures;

\item Our work is immediately deployable as it requires no modification to the program source code.

\end{itemize}

\section{Background}
In this section, we first introduce SpMV and sparse matrix storage formats, before describing the two many-core architectures targeted in
the work.

\subsection{Sparse Matrix-Vector Multiplication}
\begin{figure}[!h]
\centering
\includegraphics[width=0.50\linewidth]{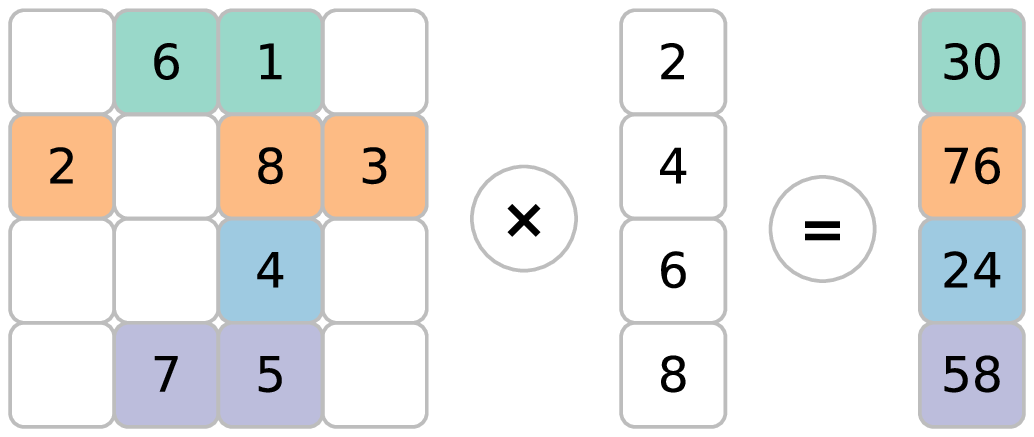}
\caption{A simple example of SpMV with a $4\times4$ matrix and a vector. The product of the SpMV is a one-dimensional vector.}
\label{fig:spmv_example}
\end{figure}

SpMV can be formally defined as $\mathbf{y}=\mathbf{Ax}$, where the input matrix, $\mathbf{A}$ ($M\times N$), is sparse, and the input,
$\mathbf{x}$ ($N\times 1$), and the output, $\mathbf{y}$ ($M\times 1$), vectors are dense. Figure~\ref{fig:spmv_example} gives a simple
example of SpMV with $M$ and $N$ equal to 4, where the number of nonzeros ($nnz$) of the input matrix is 8.


\subsection{Sparse Matrix Representation}

Since most of the elements  of a sparse matrix are zeros, it would be a waste of space and time to store these entries and perform
arithmetic operations on them. To this end, researcher have designed a number of compressed storage representations to store only the
nonzeros. We describe the sparse matrix representations targeted in this work. Note that different representations require different SpMV
implementations, and thus have different performance on distinct architectures and inputs.

\cparagraph{COO} The \textit{coordinate} (COO) format (a.k.a. IJV format) is a particularly simple storage scheme. The arrays \texttt{row},
\texttt{col}, and \texttt{data} are used to store the row indices, column indices, and values of the nonzeros. This format is a general
sparse matrix representation, because the required storage is always proportional to the number of nonzeros for any sparsity pattern.
Different from other formats,
COO stores explicitly both row indices and column indices.
Table~\ref{tbl:matrix_format} shows an example matrix in the COO format.

\begin{table}[!h]
\caption{Matrix storage formats and their data structures for the sparse matrix shown in Figure~\ref{fig:spmv_example}.}
\label{tbl:matrix_format}
\begin{center}
\scalebox{0.85}{
\begin{tabular}{cr} \toprule
\textbf{Representation} & \textbf{Specific Values}       \\
\midrule
\rowcolor{Gray} COO  &
\begin{tabular}{@{}r@{}}$row=[0, 0, 1, 1, 1, 2, 3, 3]$ \\ $col=[1, 2, 0, 2, 3, 2, 1, 2]$ \\ $data=[6, 1, 2, 8, 3, 4, 7, 5]$\end{tabular}
 \\
CSR  & \begin{tabular}{@{}r@{}}$ptr=[0, 2, 5, 6, 8]$ \\ $indices=[1, 2, 0, 2, 3, 2, 1, 2]$ \\ $data=[6, 1, 2, 8, 3, 4, 7, 5]$\end{tabular} \\
\rowcolor{Gray}
CSR5\footnote{Note that `$|$' is used to separate two data tiles.}  & \begin{tabular}{@{}r@{}}$ptr=[0, 2, 5, 6, 8]$  $tile\_ptr=[0, 1, 4]$\\
$tile\_des: bit\_flag=[T, T, F, F | T, T, T, F],$\\ $y\_off=[0, 1 | 0, 2], seg\_off=[0, 0 | 0, 0]$ \\
$indices=[1, 0, 2, 2 | 3, 1, 2, 2]$ \\  $data=[6, 2, 1, 8 | 3, 7, 4, 5]$\end{tabular}\\
 ELL  &  \begin{tabular}{@{}r@{}}$data=\begin{bmatrix}
6 & 1 & *\\
2 & 8 & 3\\
4 & * & *\\
7 & 5 & *
\end{bmatrix}$ $indices=\begin{bmatrix}
1 & 2 & *\\
0 & 2 & 3\\
2 & * & *\\
1 & 2 & *
\end{bmatrix}$ \end{tabular}
\\
\rowcolor{Gray} SELL  &  \begin{tabular}{@{}r@{}}$data=\begin{bmatrix}
6 & 1 & *\\
2 & 8 & 3\\
4 & * & \\
7 & 5 &
\end{bmatrix}$ $indices=\begin{bmatrix}
1 & 2 & *\\
0 & 2 & 3\\
2 & * & \\
1 & 2 &
\end{bmatrix}$ \\ $slices=[3, 2]$  \end{tabular}
  \\
SELL-C-$\sigma$  &  \begin{tabular}{@{}r@{}}$data=\begin{bmatrix}
2 & 8 & 3\\
6 & 1 & *\\
7 & 5 & \\
4 & * & \\
\end{bmatrix}$ $indices=\begin{bmatrix}
0 & 2 & 3\\
1 & 2 & *\\
1 & 2 & \\
2 & * & \\
\end{bmatrix}$ \\ $slices=[3, 2]$  \end{tabular}
  \\
\rowcolor{Gray} HYB  &  \begin{tabular}{@{}r@{}}ELL: $data=\begin{bmatrix}
6 & 1\\
2 & 8\\
4 & *\\
7 & 5
\end{bmatrix}$ $indices=\begin{bmatrix}
1 & 2\\
0 & 2\\
2 & *\\
1 & 2
\end{bmatrix}$ \\
COO: $row=[1]$, $col=[3]$, $data=[3]$
\end{tabular}
\\
\bottomrule
\end{tabular}}
\end{center}
\end{table}

\cparagraph{CSR} The \textit{compressed sparse row} (CSR) format is the most popular, general-purpose sparse matrix representation. This
format explicitly stores column indices and nonzeros in array \texttt{indices} and \texttt{data}, and uses a third array \texttt{ptr} to
store the starting nonzero index of each row in the sparse matrix (i.e., row pointers). For an $M\times N$ matrix, \texttt{ptr} is sized of
$M+1$ and stores the offset into the $i^{th}$ row in \texttt{ptr[i]}. Thus, the last entry of \texttt{ptr} is the total number of nonzeros.
Table~\ref{tbl:matrix_format} illustrates an example matrix represented in CSR. We see that the CSR format is a natural extension of the COO
format by using a compressed scheme. In this way, CSR can reduce the storage requirement. More importantly, the introduced \texttt{ptr}
facilitates a fast query of matrix values and other interesting quantities such as the number of nonzeros in a particular row.

\cparagraph{CSR5} To achieve near-optimal load balance for matrices with any sparsity structures, CSR5 first evenly partitions all nonzero
entries to multiple 2D tiles of the same size. Thus when executing parallel SpMV operation, a compute core can consume one or more 2D
tiles, and each SIMD lane of the core can deal with one column of a tile. Then the main skeleton of the CSR5 format is simply a group of 2D
tiles. The CSR5 format has two tuning parameters: $\omega$ and $\sigma$, where $\omega$ is a tile's width and $\sigma$ is its height. CSR5
is an extension to the CSR format~\cite{DBLP:conf/ics/0002V15}. Apart from the three data structures from CSR, CSR5 introduces another two
data structures: a tile pointer \texttt{tile\_ptr} and a tile descriptor \texttt{tile\_des}. Table~\ref{tbl:matrix_format} illustrates an
example matrix represented in CSR5, where $\omega$=$\sigma$=$2$.

\cparagraph{ELL} The \texttt{ELLPACK} (ELL) format is suitable for the vector architectures. For an $M\times N$ matrix with a maximum of
$K$ nonzeros per row, ELL stores the sparse matrix in a dense $M\times K$ array (\texttt{data}), where the rows having fewer than $K$ are
padded. Another data structure \texttt{indices} stores the column indices and is zero-padded in the same way with that of \texttt{data}.
Table~\ref{tbl:matrix_format} shows the ELL representation of the example sparse matrix, where $K=3$ and the data structures are padded
with \texttt{*}.
The ELL format would waste a decent amount of storage.
To mitigate this issue, we can combine ELL with another general-purpose format such as CSR or COO (see Section~\ref{sec:hyb}).

\cparagraph{SELL and SELL-C-$\sigma$} \textit{Sliced} ELL (SELL) is an extension to the ELL format by partitioning the input matrix into
strips of $C$ adjacent rows~\cite{DBLP:conf/hipeac/MonakovLA10}. Each strip is stored in the ELL format, and the number of nonzeros stored
in ELL may differ over strips. Thus, a data structure \texttt{slice} is used to keep the strip information. Table~\ref{tbl:matrix_format}
demonstrates a matrix represented in the SELL format when $C=2$. A variant to SELL is the SELL-C-$\sigma$ format which introduces sorting
to save storage overhead~\cite{DBLP:journals/siamsc/KreutzerHWFB14}. That is, they choose to sort the matrix rows not globally but within
$\sigma$ consecutive rows. Typically, the sorting scope $\sigma$ is selected to be a multiple of $C$. The effect of local sorting is shown
in Table~\ref{tbl:matrix_format} with $C=2$ and $\sigma =4$.

\cparagraph{HYB} \label{sec:hyb} The HYB format is a combination of ELL and COO, and it stores the majority of matrix nonzeros in ELL while
the remaining entries in COO~\cite{DBLP:conf/sc/BellG09}. Typically, HYB stores the \textit{typical} number of nonzeros per row in the ELL
format and the \textit{exceptionally} long rows in the COO format. In the general case, this \textit{typical} number ($K$) can be
calculated directly from the input matrix. Table~\ref{tbl:matrix_format} shows an example matrix in this hybrid format, with $K=2$.


\section{Evaluation Setup}

\subsection{Hardware Platforms}
Our work targets two many-core architectures designed for HPC, described as follows.

\begin{figure}[!t]
\centering
\includegraphics[width=0.50\textwidth]{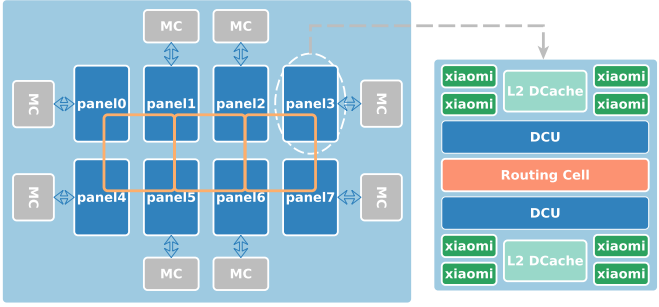}
\caption{A high-level view of the FT-2000Plus architecture. Processor cores are groups into panels (left) where each panel contains eight ARMv8
based Xiaomi cores (right).} \label{fig:ft2000_architecture}
\end{figure}

\cparagraph{FT-2000Plus} Figure~\ref{fig:ft2000_architecture} gives a high-level view of the FT-2000Plus architecture.  This
architecture~\cite{ft2000} integrates 64 ARMv8 based Xiaomi cores, offering a peak performance of 512 Gflops for double-precision
operations, with a maximum power consumption of 100 Watts. The cores can run up to 2.4 GHz, and are groups into eight panels with eight
cores per panel. Each core has a private L1 data cache of 32KB, and a dedicated 2MB L2 cache shared among four cores. The panels are connected through two directory control units (\texttt{DCU}) and a routing
cell~\cite{DBLP:conf/hotchips/Zhang15}, where cores and caches are linked via a 2D mesh network. External I/O are managed by the DDR4
memory controllers (\texttt{MC}), and the routing cells at each panel link the \texttt{MCs} to the \texttt{DCUs}.


\cparagraph{Intel KNL} A KNL processor has a peak performance of 6 Tflops/s and 3Tflops/s respectively for single- and double-precision
operations~\cite{DBLP:conf/ipps/RamosH17}. A KNL socket can have up to 72 cores where each core has four threads running at 1.3 GHz. Each
KNL core has a private L1 data and a private L1 instruction caches of 32KB, as well as two vector processor units (VPU), which differs from the KNC core~\cite{DBLP:conf/wosp/FangSZXCV14, DBLP:journals/corr/FangVSZCX13}. 

\begin{figure}[!t]
\centering
\includegraphics[width=0.50\textwidth]{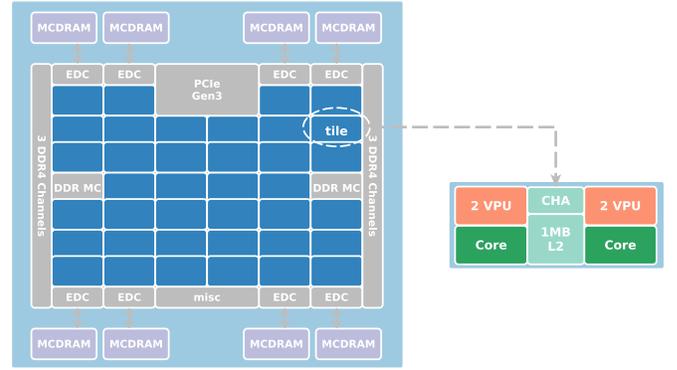}
\caption{A high-level overview of the Intel KNL architecture. Cores are grouped into tiles (left) with two cores per title (right).}
\label{fig:knl_architecture}
\end{figure}

As shown in Figure~\ref{fig:knl_architecture}, KNL cores are organized around 36 \texttt{tiles} where each title has two cores. Each title
also has a private, coherent 1MB L2 data cache shared among cores, which is managed by the cache/home agent (CHA). Tiles are connected into
a 2D mesh to facilitate coherence among the distributed L2 caches. Furthermore, a KNL chip has a `near' and a `far' memory components. The
‘near’ memory components are multi-channel DRAM (MCDRAM) which are connected to the tiles through the MCDRAM Controllers (EDC). The `far'
memory components are DDR4 RAM connected to the chip via the DDR memory controllers (DDR MC). While the `near' memory is smaller than the
`far' memory, it provides 5x more bandwidth over traditional DDRs. Depending how the chip is configured, some parts or the entire near
memory can share a global memory space with the far memory, or be used as a cache.
In this context, MCDRAM is used in the cache mode and the dataset can be hold in the high-speed memory.

\subsection{Systems Software}
Both platforms run a customized Linux operating system with Kernel v4.4.0 on FTP and v3.10.0 on KNL. For compilation, we use gcc v6.4.0 on
FTP and Intel icc v17.0.4 on KNL with the default ``-O3" compiler option. We use the OpenMP threading model on both platforms with 64
threads on FTP and 272 threads on KNL.

\subsection{Datasets}
Our experiments use a set of 956 square matrices (with a total size of 90 GB) from the SuiteSparse matrix
collection~\cite{DBLP:journals/toms/DavisH11}. The number of nonzeros of these matrices ranges from 100K to 20M. The dataset includes both
regular and irregular matrices, covering application domains ranging from scientific computing to social networks.

\subsection{SpMV Implementation} Algorithm~\ref{alg:spmv_bench} illustrates our library-based SpMV implementation using the CSR5 format as
an example. Our library takes in
 the raw data of the \texttt{Matrix Market} format into memory of the COO format. Then we convert the COO-based data into our target
storage format (CSR, CSR5, ELL, SELL, or HYB). When calculating SpMV, we use the OpenMP threading model for parallelization.
This process
is format dependent, i.e., the basic task can be a row, a block row, or a data tile.
When calculating a single element
of $\mathbf{y}$ falls into different tasks, we will have to gather the partial results.
This efficient data gathering can be achieved by manually vectorize the SpMV code with \texttt{intrinsics}.
Due to the lack of the \texttt{gather/scatter} function, we do not
use the \texttt{neon} intrinsics on FTP. This is because our experimental results show that explicitly using the intrinsics results in a
loss in performance, compared with the C version. When measuring the performance, we run the experiments for $FRQ$ times and calculate
the mean results.

\begin{algorithm}[t]
\caption{The SpMV Bench based on CSR5}\label{alg:spmv_bench}
\begin{algorithmic}[1]
\Procedure {BenchSpMV}{$\mathbf{A}$, $\mathbf{x}$; $\mathbf{y}$} \LState COOFMT* $ctx$ $\leftarrow$ ReadMatrix($mtx\_file$) \LState
CSR5FMT* $ntx$ $\leftarrow$ ConvertToCSR5($ctx$) \For{$t \leftarrow 1, 2, \ldots, FRQ$}
	\LState \texttt{\#pragma omp for}
	\For{each tile $dt$ in $ntx$}
		\LState $y'$ $\leftarrow$ CalculateSpMV($dt$)
	\EndFor
	\LState UpdateProduct($y$, $y'$)
\EndFor \LState DumpInfo($runT$, $gflops$, $bw$) \LState DeleteMTX($ctx$, $ntx$) \EndProcedure
\end{algorithmic}
\end{algorithm}



\section{Memory Allocation and Code Vectorization}

SpMV performance depends on a number of factors on a many-core architecture. These include memory allocation and code optimization
strategies, and the sparse matrix representation. The focus of this work is to identify the optimal sparse matrix representation that leads
to the quickest running time. To isolate the problem, we need to ensure that performance comparisons of different representations are
conducted on the best possible memory allocation and code optimization strategies. To this end, we investigate how Non-Uniform Memory
Access (NUMA) and code vectorization affect the SpMV performance on FTP and KNL. We then conduct our experiments on the best-found strategy
of NUMA memory allocation and vectorization.


\subsection{The Impact of NUMA Bindings}
Unlike the default setting of KNL, the FTP architecture exposes multiple NUMA nodes where a group of eight cores are directly connected to a local
memory module. Indirect access to remote memory modules is possible but slow. This experiment evaluates the impact of NUMA on SpMV
performance on FTP. We use the Linux NUMA utility, \texttt{numactl}, to allocate the required data buffers from the local memory module of
a running processor.

\begin{figure}[!t]
\centering
\includegraphics[width=0.42\textwidth]{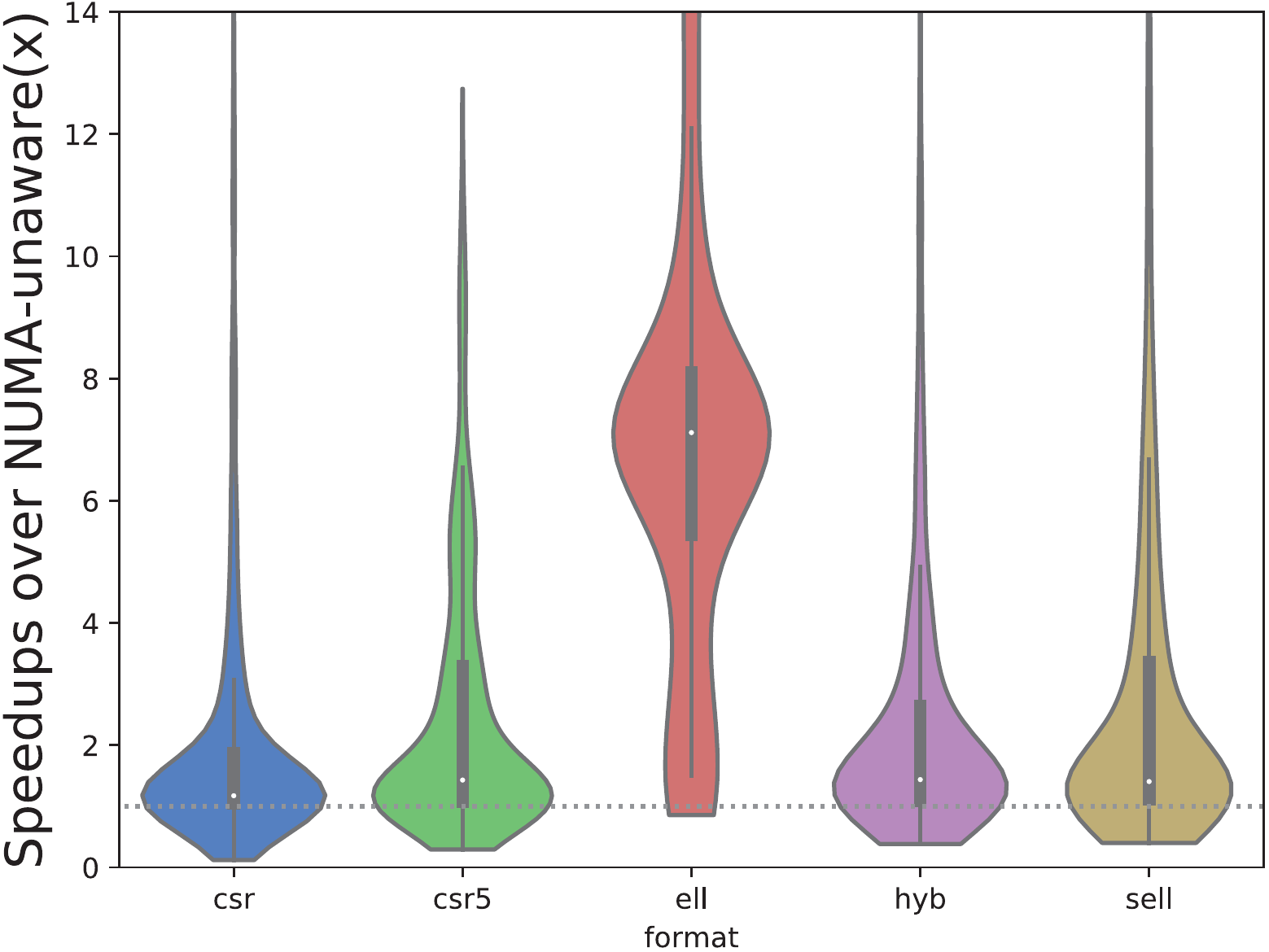}
\caption{The violin diagram shows the speedup distribution of NUMA-aware memory allocation over NUMA-unaware memory allocation on FTP.
The thick black line shows where 50\% of the data lie. NUMA-aware memory allocation can significantly improve the SpMV
performance.}
\label{fig:numactl_perf}
\end{figure}

Figure~\ref{fig:numactl_perf} show the performance improvement when using NUMA-aware over non-NUMA-aware memory allocation across five
sparse matrix representation. We see that static NUMA bindings enables significant performance gains for all the five storage formats on
FTP. Compared with the case without tuning, using the NUMA tunings can yield an average speedup of 1.5x, 1.9x, 6.0x, 2.0x, and 1.9x for
CSR, CSR5, ELL, SELL, and HYB, respectively. Note that we have achieved the maximum speedup for the ELL-based SPMV. This is due to the fact
that using ELL allocates the largest amount of memory buffers, and the manual NUMA tunings can ensure that each NUMA node accesses its
local memory as much as possible.

\subsection{The Impact of Vectorization on KNL} \label{sec:vec_effect}
\begin{figure}[!t]
\centering
\includegraphics[width=0.36\textwidth]{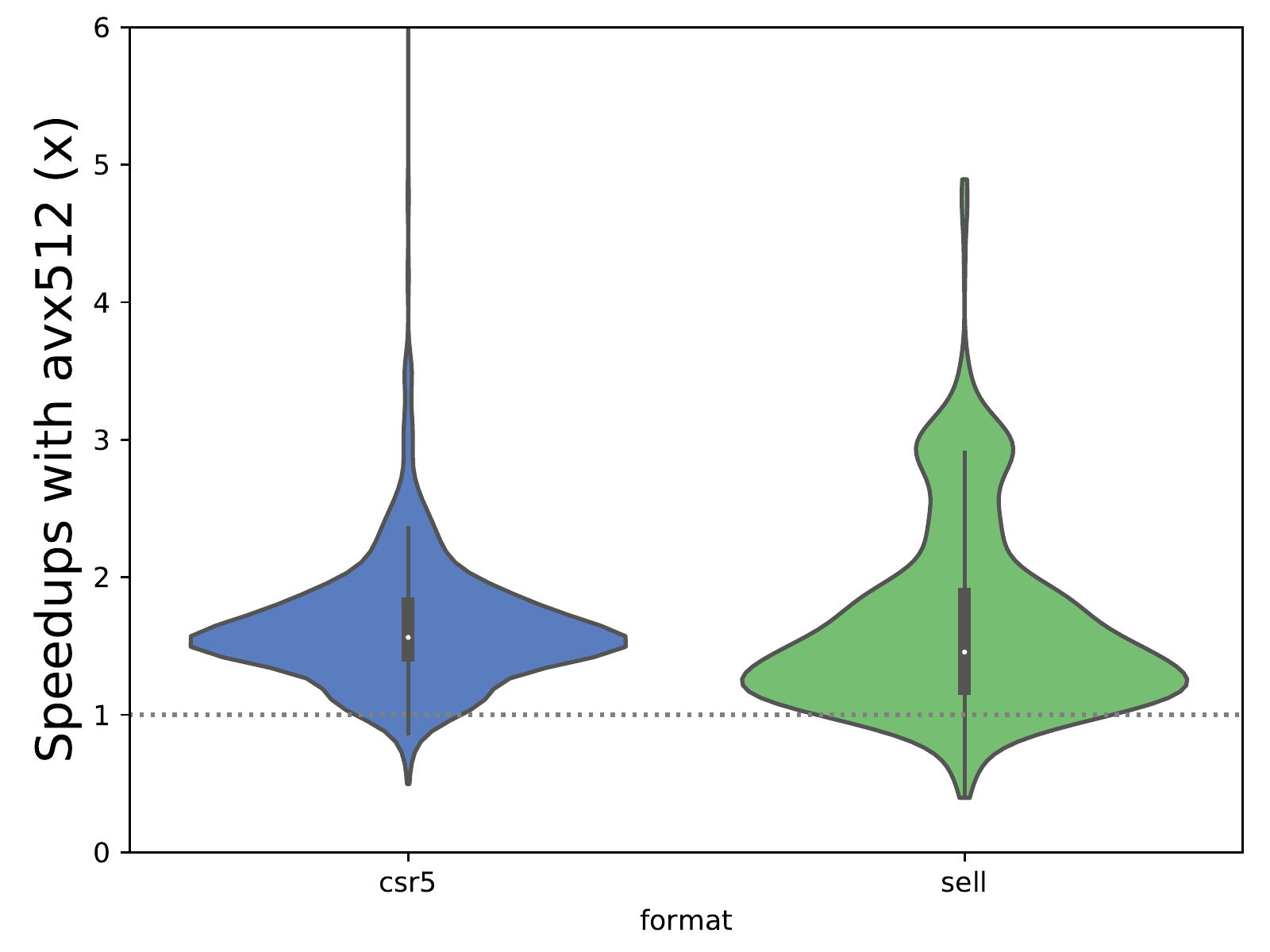}
\caption{The SpMV performance with and without explicit vectorization on KNL. }
\label{fig:avx512_perf}
\end{figure}

The two many-core architectures considered in the work support SIMD vectorization~\cite{DBLP:journals/concurrency/FangVLS15}. KNL and FTP have a vector unit of 512 and 128 bits
respectively. Figure~\ref{fig:avx512_perf} shows that vectorization performance of CSR5 and SELL-based SpMV on KNL. Overall, we see that
manually vectorizing the code using vectorization \texttt{intrinsics} can significantly improve the SpMV performance on KNL. Compared with
the code without manual vectorization, the vectorized code yields a speedup of 1.6x for CSR5 and 1.5x for SELL. However, we observe no
speedup for vectorized code on FTP. This is that because FTP has no support of the \texttt{gather} operation which is essential for
accessing elements from different locations of a vector. By contrast, KNL supports \texttt{\_mm512\_i32logather\_pd}, which improves the
speed of the data loading process. Therefore, for the remaining experiments conducted in this work, we manually vectorize the code on KNL
but not on FTP.

\subsection{FTP versus KNL}
\begin{figure}[!t]
\centering
\includegraphics[width=0.40\textwidth]{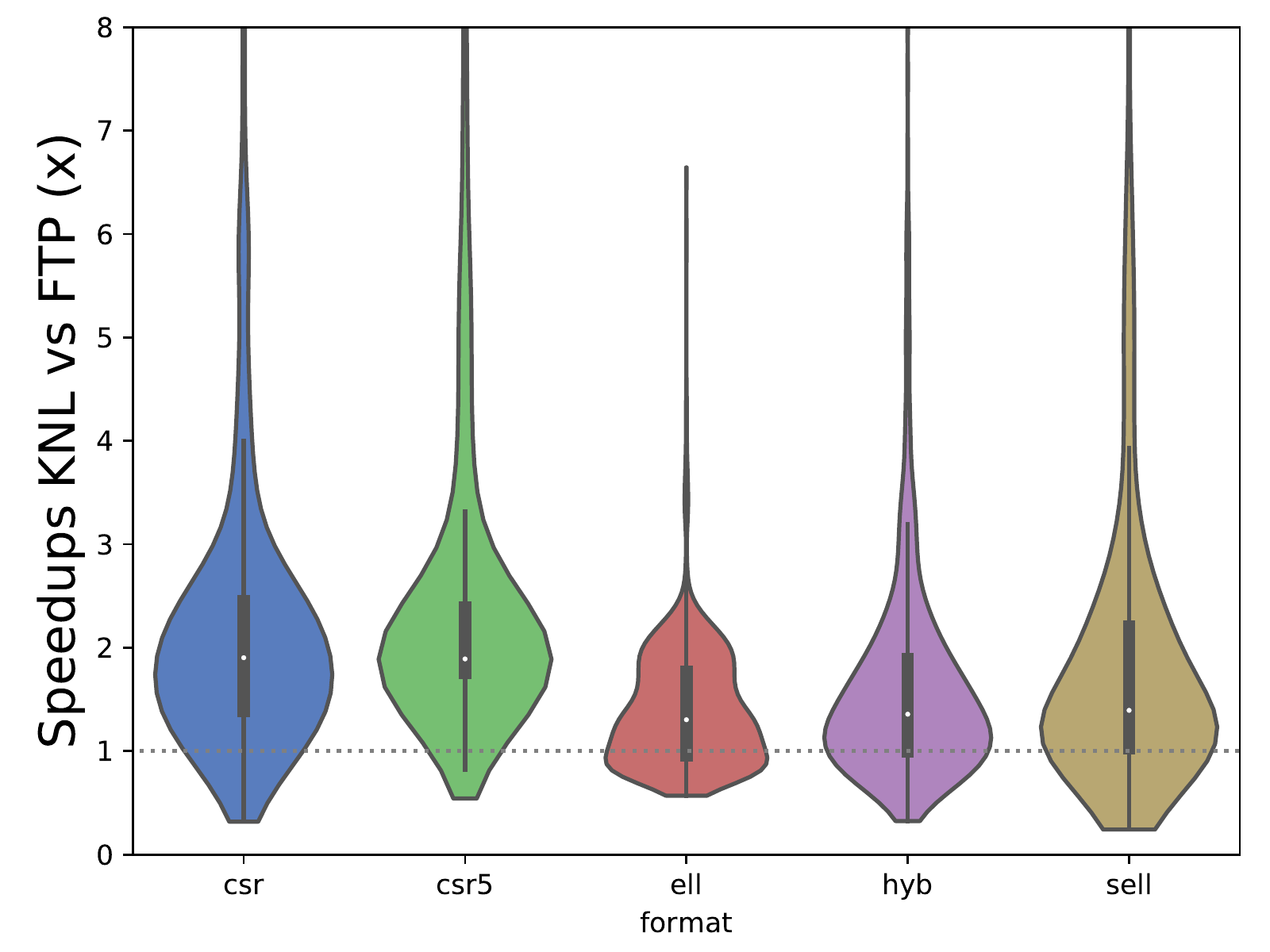}
\caption{Comparing the SpMV performance between KNL and FTP.
The x-axis labels different sparse matrix representation,
and the y-axis denotes the achieved speedup of KNL over FTP. }
\label{fig:knl_vs_ft_sp}
\end{figure}

Figure~\ref{fig:knl_vs_ft_sp} shows the performance comparison between KNL and FTP. In general, we observe that SpMV on KNL runs faster
than it on FTP for each format. The average speedup of KNL over FTP is 1.9x for CSR, 2.3x for CSR5, 1.3x for ELL, 1.5x for SELL, and 1.4x
for HYB. The performance disparity comes from the difference in the memory hierarchy of the architectures. KNL differs from FTP in that it
has a high-speed memory, a.k.a., MCDRAM, between the L2 cache and the DDR4 memory. MCDRAM can provide 5x more memory bandwidth over the
traditional DDR memory. Once the working data is loaded into this high-speed memory, the application can then access the data with a higher
memory bandwidth which leads to a better overall performance. SpMV on KNL also benefits from the support of \texttt{gather/scatter}
 operations (see Section~\ref{sec:vec_effect}). This is
key for the overall SpMV performance, which is limited by the scattered assess of the input vector.
To sum up, we would have a significant performance increase when the aforementioned memory features are enabled on FTP.

From Figure~\ref{fig:knl_vs_ft_sp}, we also observe FTP outperforms KNL on some matrices, specially when the size of the input matrices is
small. The performance disparity is due to the fact that KNL and FTP differ in L2 cache in terms of both capacity and coherence protocol.


\begin{table}[!t]
\centering
\caption{The best format distribution on FTP and KNL.}
\label{tbl:format_distribution}
\scalebox{0.80}{
\begin{tabular}{@{}llllll@{}}
\toprule
    & \textbf{CSR}         & \textbf{CSR5}        & \textbf{ELL}       & \textbf{SELL}        & \textbf{HYB}         \\ \midrule
\rowcolor{Gray} FTP & 127(14.1\%) & 149(16.5\%) & 22(2.4\%) & 443(49.2\%) & 160(17.6\%) \\
KNL & 493(51.8\%) & 273(28.7\%) & 39(4.1\%) & 121(12.7\%) & 25(2.6\%)   \\ \bottomrule
\end{tabular}}
\end{table}

\begin{figure}[!t]
\centering
\subfigure[On FTP with 64 threads]{\label{fig:perf_on_ftp}\includegraphics[width=0.50\textwidth]{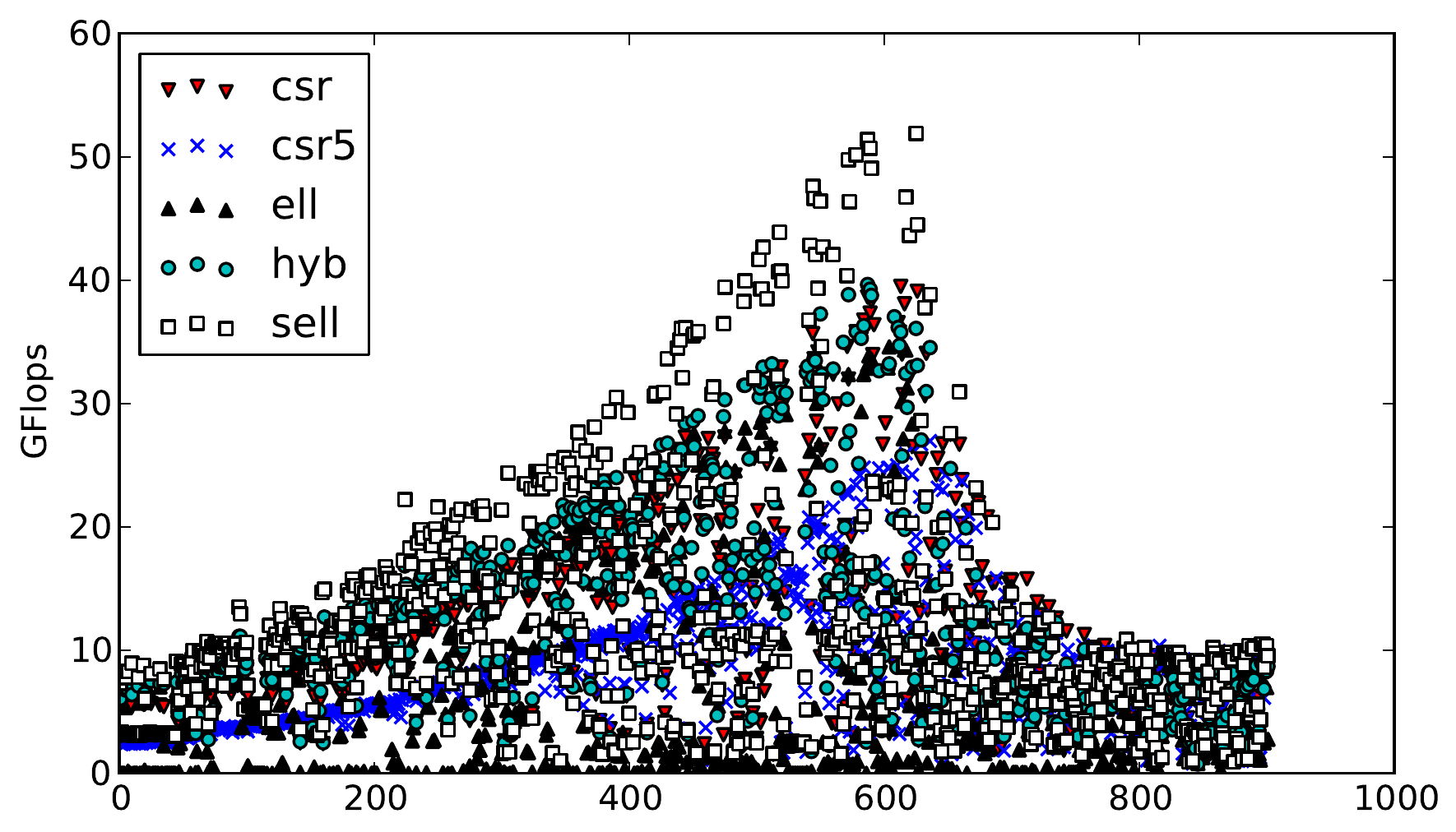}}
\subfigure[On KNL with 272 threads]{\label{fig:perf_on_knl}\includegraphics[width=0.50\textwidth]{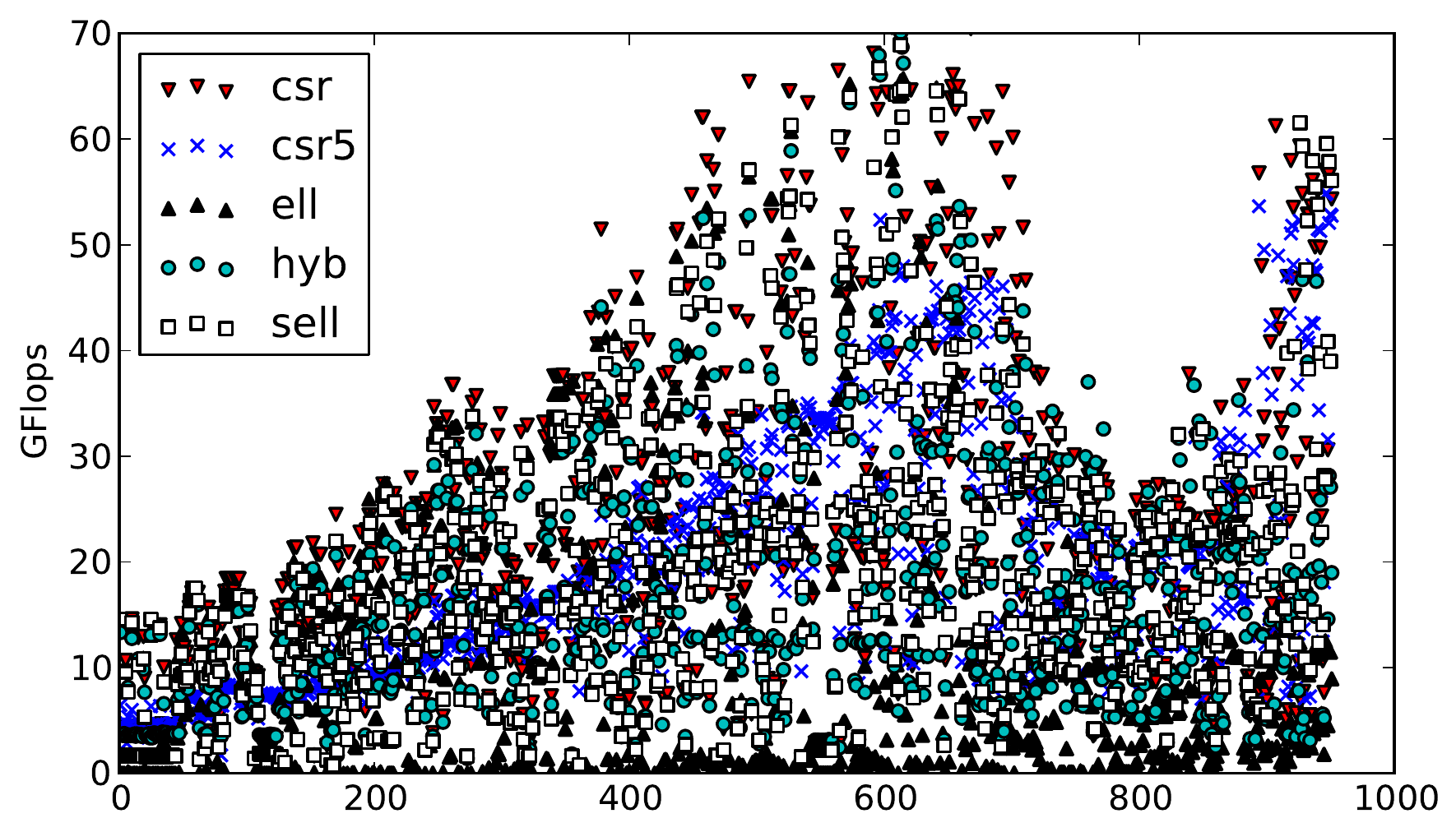}}
\caption{The overall performance of SpMV on FTP and KNL.
The x-axis labels different sparse matrices ordered by the number of nonzeros,
and the y-axis denotes the achieved SpMV performance in GFlops. }
\label{fig:perf_spmv}
\end{figure}

\begin{table}[!t]
\centering
\caption{The average slowdowns over all the matrices when using a single format instead of the individual best.}
\label{tbl:perf_slowdowns}
\scalebox{0.80}{
\begin{tabular}{@{}llllll@{}}
\toprule
    & \textbf{CSR}         & \textbf{CSR5}        & \textbf{ELL}       & \textbf{SELL}        & \textbf{HYB}         \\ \midrule
\rowcolor{Gray} FTP & 1.4x 		  & 1.8x			& 6.4x		& 1.3x		  & 1.3x			\\
KNL & 1.3x		  & 1.4x 			& 8.7x		& 1.5x		  & 1.6x   \\ \bottomrule
\end{tabular}}
\end{table}
\subsection{Optimal Sparse Matrix Formats}
Figure~\ref{fig:perf_spmv} shows the overall performance of SpMV on FTP and KNL . We see that there is no ``one-size-fits-all'' format
across inputs and architectures. On the FTP platform, SELL is the optimal format for around 50\% of the sparse matrices and ELL gives the
worse performance on most of the cases. On the KNL platform, CSR gives the best performance for most of the cases, which is followed by
CSR5 and SELL. On KNL, ELL and HYB give the best performance for just a total of 64 sparse matrices (Table~\ref{tbl:format_distribution}).


Table~\ref{tbl:perf_slowdowns} shows the average slowdowns when using a fixed format across test cases over the optimal one. The slowdown
has a negative correlation with how often a given format being optimal. For example, CSR gives the best overall performance on KNL and as
such it has the lowest overall slowdown. Furthermore, SELL and HYB have a similar average slowdown on FTP because they often deliver
similar performance (see Figure~\ref{fig:perf_on_ftp}).

Given that the optimal sparse matrix storage format varies across architectures and inputs, finding the optimal format is a non-trivial
task. What we like to have is an adaptive scheme that can automatically choose the right format for a given input and architecture. In the
next section, we describe how to develop such a scheme using machine learning.

\begin{figure}[t!]
  \centering
  \includegraphics[width=0.5\textwidth]{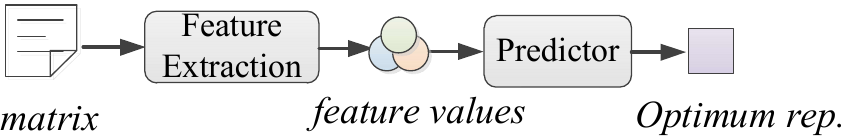}\\
  \caption{Overview of our approach.}\label{fig:overview}
\end{figure}
\section{Adaptive Representation Selection}
\subsection{Overall Methodology}
Our approach takes a \emph{new}, \emph{unseen} sparse matrix and is able to predict the optimal or near optimal sparse matrix
representation for a given architecture. An overview of our approach can be seen in Figure~\ref{fig:overview}, and is described in more
details in Section~\ref{sec:modelling}. Our predictive model is built upon the \texttt{scikit-learn} machine learning
package~\cite{scikit-learn}.

For a given sparse matrix, our approach will collect a set of information, or \emph{features}, to capture the characteristics of the
matrix. The set of feature values can be collected at compile time or during runtime. Table~\ref{tbl:features} presents a full list of all our
considered features. After collecting the feature values, a machine learning based predictor (that is trained offline) takes in the feature
values and predicts which matrix representation should be used for the target architecture. We then transform the matrix to the predicted
format and generate the computation code for that format.

\subsection{Predictive Modeling}
\label{sec:modelling} Our model for predicting the best sparse matrix representation is a decision tree model. We have evaluated a number
of alternate modelling techniques, including regression, Naive Bayes and K-Nearest neighbour (see also Section~\ref{sec:alter}).
We chose the decision tree model because it gives the best performance and can be easily interpreted compared to other
black-box models. The input to our model is a set of features extracted from the input matrix. The output of our model is a label that
indicates which sparse matrix representation to use.

Building and using such a model follows the 3-step process for supervised machine learning: (i) generate training data (ii) train a
predictive model (iii) use the predictor, described as follows.

\begin{figure}[t!]
  \centering
  \includegraphics[width=0.5\textwidth]{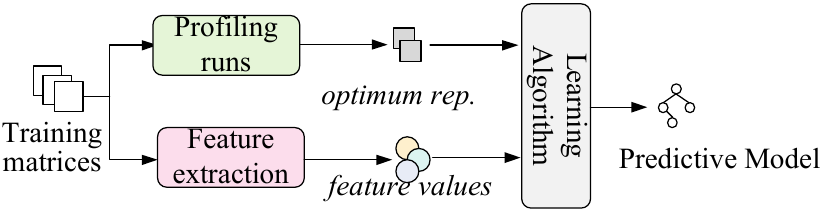}
  \caption{The training process of our predictive model.}
  \label{fig:training}
\end{figure}

\subsubsection{Training the Predictor}
Our method for training the predictive model is shown in Figure~\ref{fig:training}.
We use the same approach to train a model for each
targeting architecture. To train a predictor we first need to find the best sparse matrix representation for each of our training matrix
examples, and extract features. We then use this set of data and classification labels to train our predictor model.

\cparagraph{Generating Training Data} We use five-fold-cross validation for training. This standard machine learning technique works by
selecting 20\% samples for testing and using 80\% samples for training. To generate the training data for our model we used 756
sparse matrices from the SuiteSparse matrix collection. We execute SpMV using each sparse matrix representation a number of times until the gap of the
upper and lower confidence bounds is smaller than 5\% under a 95\% confidence interval setting. We then record the best-performing matrix
representation for each training sample on both KNL and FTP. Finally, we extract the values of our selected set of features from each matrix.

\cparagraph{Building The Model} The optimal matrix representation labels, along with their corresponding feature set, are passed to our
supervised learning algorithm. The learning algorithm tries to find a correlation between the feature values and optimal representation
labels. The output of our learning algorithm is a version of our decision-tree based model. Because we target two platforms in this paper,
we have constructed two predictive models, one model per platform. Since training is performed off-line and only need to be carried out
once for a given architecture, this is a \emph{one-off} cost.

\cparagraph{Total Training Time} The total training time of our model is comprised of two parts: gathering the training data, and then
building the model. Gathering the training data consumes most of the total training time, in this paper it took around 3 days for the two
platforms. In comparison actually building the model took a negligible amount of time, less than 10 ms.

\subsubsection{Features} \label{features}
One of the key aspects in building a successful predictor is developing the right features in order to characterize the input. Our
predictive model is based exclusively on static features of the target matrix and no dynamic profiling is required. Therefore, it can be
applied to any hardware platform. The features are extracted using our own \texttt{Python} script. Since our goal is to develop a \emph{portable},
architecture-independent approach, we do not use any hardware-specific features~\cite{DBLP:conf/cse/FangVS11}.

\cparagraph{Feature Selection}
We considered a total of seven candidate raw features (Table~\ref{tbl:features}) in this work.
Some features were chosen from our intuition based on factors that can affect SpMV performance e.g. \textit{nnz\_frac} and \textit{variation},
other features were chosen based on previous work~\cite{DBLP:conf/ics/SedaghatiMPPS15}.
Altogether, our candidate features should be able to represent the intrinsic parts of a SpMV kernel.

\begin{table}[!t]
\centering
\caption{The features used in our model.}
\label{tbl:features}
\small
\begin{tabular}{@{}ll@{}}
\toprule
\textbf{Features}   & \textbf{Description}                                 \\ \midrule
\rowcolor{Gray} n\_rows   & number of rows                          \\
n\_cols   & number of columns                       \\
\rowcolor{Gray} nnz\_frac & percentage of nonzeros                  \\
nnz\_min  & minimum number of nonzeros per row      \\
\rowcolor{Gray} nnz\_max  & maximum number of nonzeros per row      \\
nnz\_avg  & average number of nonzeros per row      \\
\rowcolor{Gray} nnz\_std  & standard derivation of nonzeros per row \\
variation  & matrix regularity \\ \bottomrule
\end{tabular}
\end{table}

\cparagraph{Feature Scaling} The final step before passing our features to a machine learning model is scaling each scalar value of the
feature vector to a common range (between 0 and 1) in order to prevent the range of any single feature being a factor in its importance.
Scaling features does not affect the distribution or variance of their values. To scale the features of a new image during deployment we
record the minimum and maximum values of each feature in the training dataset, and use these to scale the corresponding features.

\subsubsection{Runtime Deployment}
Deployment of our predictive model is designed to be simple and easy to use. To this end, our approach is implemented as an API. The API
has encapsulated all of the inner workings, such as feature extraction and matrix format translation. We also provide a source to source
translation tool to transform the computation of a given SpMV kernel to each of target representations. Using the prediction label, a
runtime can choose which SpMV kernel to use.

\subsection{Predictive Model Evaluation}
We use cross-validation to evaluate our approach. Specially, we randomly split the 965 matrices into two parts: 756 matrices for training
and 200
 matrices for testing. We learn a model with the training matrices and five representations.  We then evaluate the learned model by applying it to make prediction on the 200
testing matrices. We repeat this process multiple times to ensure each matrix is tested at least once.

\begin{figure}[!t]
\centering
\subfigure[FTP]{\label{fig:perf_on_ftp_predict}\includegraphics[width=0.36\textwidth]{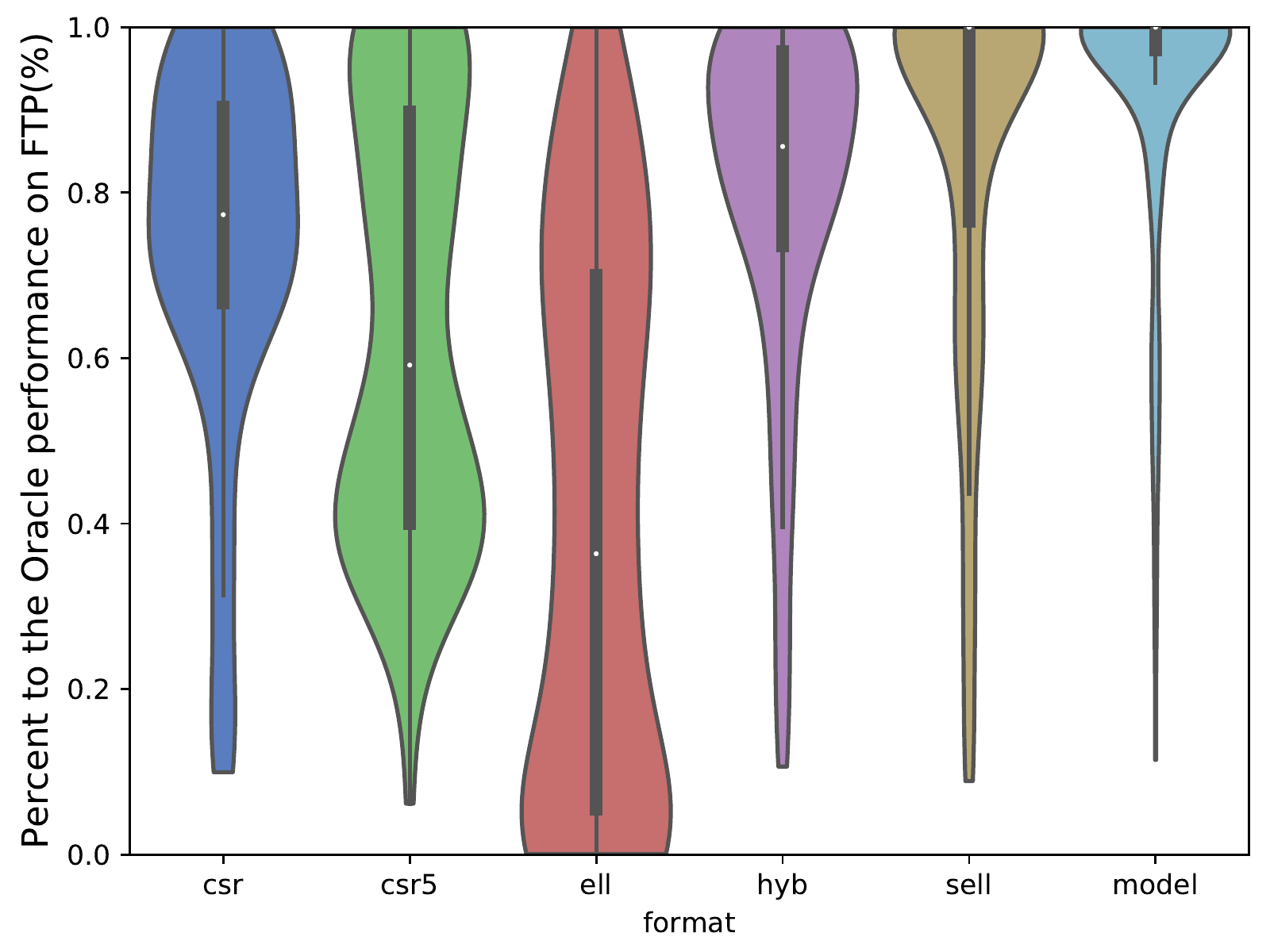}}
\subfigure[KNL]{\label{fig:perf_on_knl_predict}\includegraphics[width=0.36\textwidth]{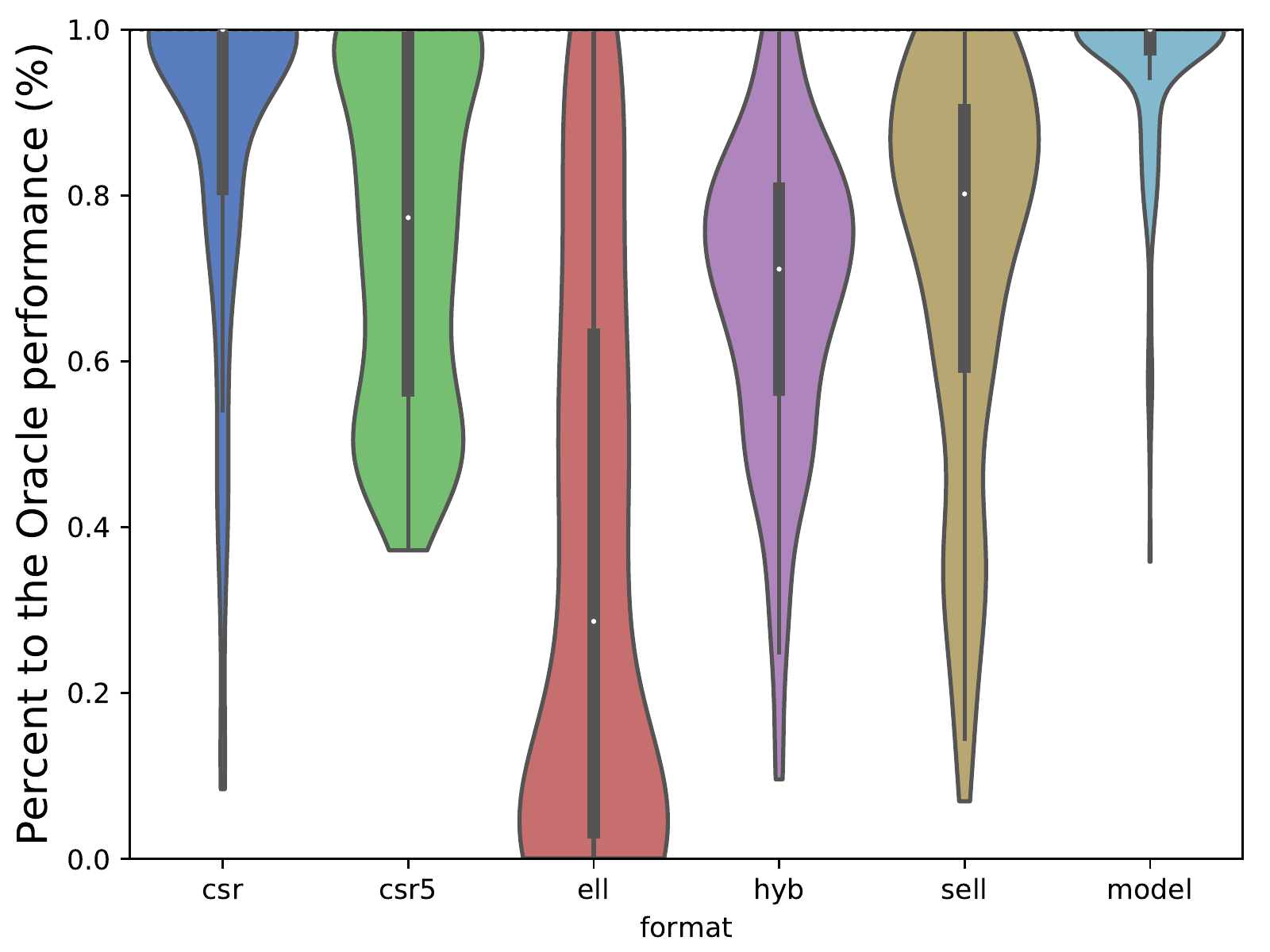}}
\caption{The predicted performance of SpMV on FTP and KNL.
We show the achieved SpMV performance with respect to the best available performance across sparse matrix format. }
\label{fig:perf_predict}
\end{figure}

Figure~\ref{fig:perf_predict} shows that our predictor achieves, on average, 91\% and 95\% of the best available SpMV performance (found
through exhaustive search) on FTP and KNL respectively. It outperforms a strategy that uses only a single format. As we have analyzed in
Table~\ref{tbl:perf_slowdowns}, SELL and HYB can achieve a better performance than the other three formats on FTP. But they are still
overtaken by our predictor. On KNL, however, the performance of our predictor is followed by using the CSR representation. Also, we note
that using the ELL representation yields poor performance on both FTP and KNL. This shows that our predictor is highly effective in
choosing the right sparse matrix representation.

\begin{figure}[t!]
  \centering
  \includegraphics[width=0.5\textwidth]{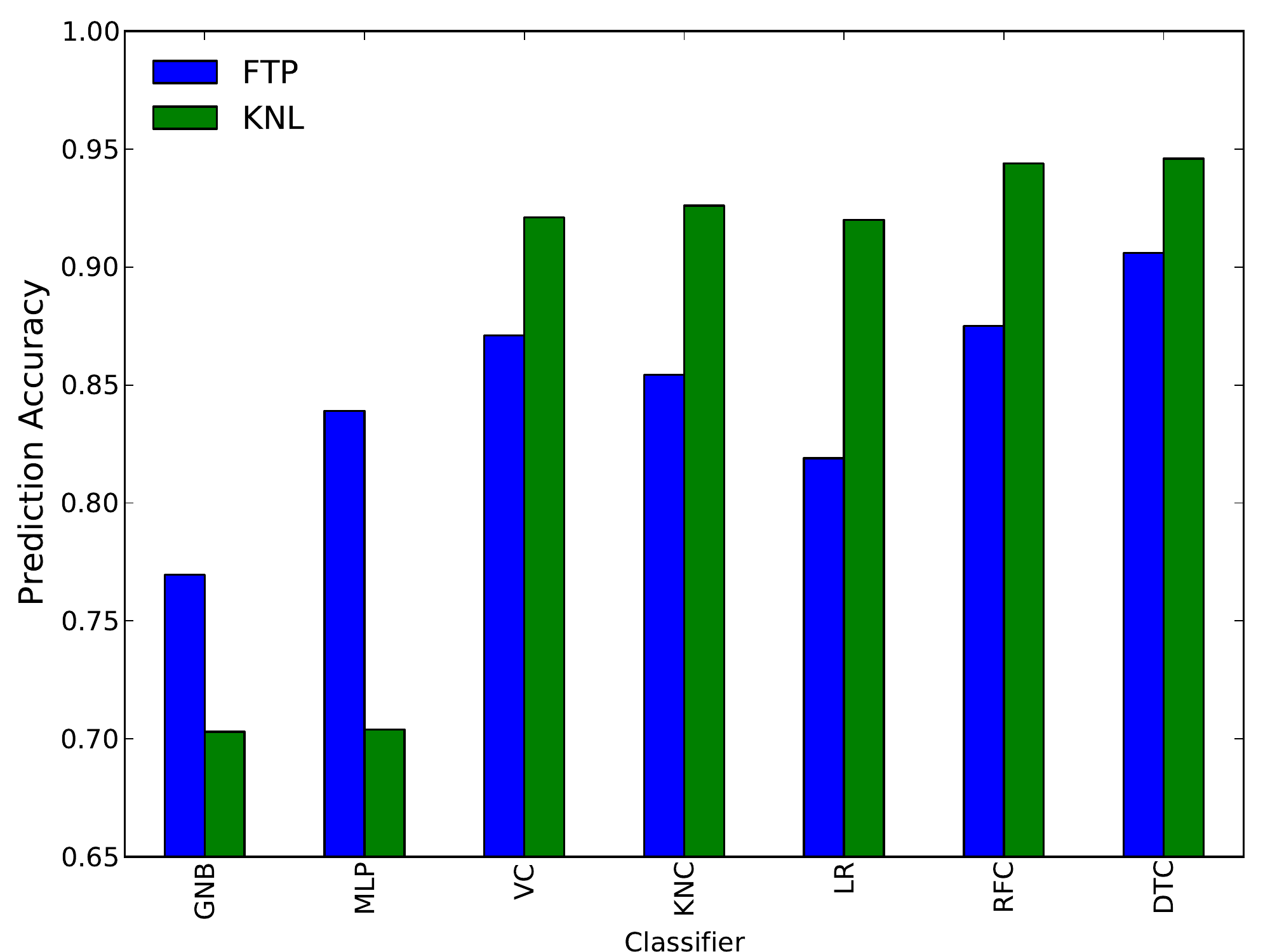}
  \caption{Compare to alternative modeling techniques}\label{fig:alternativm}
\end{figure}

\subsection{Alternative Modeling Techniques \label{sec:alter}}
Figure~\ref{fig:alternativm} shows resulting performance with respect to the best available performance when using different techniques to
construct the predictive model.  In addition to our decision tree based model (DTC), we also consider Gaussian na\"{\i}ve bayes (GNB),
multilayer perception (MLP), soft voting/majority rule Classification (VC), k-Nearest Neighbor (KNC, k=1), logistic regression (LR), random
forests classification (RFC). Thanks to the high-quality features, all classifiers are highly accurate in choosing sparse matrix
representation. We choose DTC because its accuracy is comparable to alternative techniques and can be easily visualized and interpreted.

\section{Related Work}
SpMV has been extensively studied on various platforms over the past few decades~\cite{DBLP:journals/ijhpca/Mellor-CrummeyG04,
DBLP:conf/sc/PinarH99, DBLP:conf/sc/WilliamsOVSYD07}. A large body of work has been dedicated to efficient and scalable SpMV on multi-core
and many-core processors. However, our work is the first to conduct a comprehensive study on KNL and FTP.

\cparagraph{On the SIMD processors}, some researchers have designed new compressed formats to enable efficient
SpMV~\cite{DBLP:journals/pc/WilliamsOVSYD09, DBLP:conf/sc/BellG09, DBLP:journals/tjs/GoumasKAKK09, DBLP:journals/ijhpca/ImYV04,
DBLP:conf/sc/WilliamsOVSYD07}. Liu et al. propose a storage format CSR5~\cite{DBLP:conf/ics/0002V15}, which is a tile-based format. CSR5
offers high-throughput SpMV on various platforms and shows good performance for both regular and irregular matrices. And the format
conversion from CSR to CSR5 can be as low as the cost of a few SpMV operations. On KNC, Liu et al. have identified and addressed several
bottlenecks~\cite{DBLP:conf/ics/LiuSCD13}. They exploit the salient architecture features of KNC, use specialized data structures with
careful load balancing to obtain satisfactory performance.
Wai Teng Tang et al. propose a SpMV format called vectorized hybrid COO+CSR (VHCC)~\cite{DBLP:conf/cgo/TangZLLHLG15},
which employs a 2D jagged partitioning, tiling and vectorized prefix sum computations to improve hardware resource.
Their SpMV implementation achieves an average 3x speedup over Intel MKL for a range of scale-free matrices.

In recent years, ELLPACK is the most successful format on the wide SIMD processors.
Bell and Garland propose the first ELLPACK-based format HYB, combining ELLPACK with COO formats~\cite{DBLP:conf/sc/BellG09}.
The HYB can improve the SpMV performance especially for matrix which are ``wide''.
Sliced ELLPACK format has been proposed by Monakov et al., where slices of the matrix are packed in ELLPACK format separately~\cite{DBLP:conf/hipeac/MonakovLA10}.
BELLPACK is a format derived from ELLPACK, which sorts rows of the matrix by the number of nonzeros per row~\cite{DBLP:conf/ppopp/ChoiSV10}.
Monritz Kreutzer et al. have designed a variant of Sliced ELLPACK SELL-C-sigma based on Sliced ELLPACK as a SIMD-friendly data format, in which slices are sorted~\cite{DBLP:journals/siamsc/KreutzerHWFB14}.

There have also been studies on optimizing SpMV dedicated for \textbf{SIMT GPUs}~\cite{DBLP:conf/iccS/MaggioniB13, DBLP:conf/sc/BellG09, DBLP:conf/cgo/VenkatSHS14, DBLP:conf/cgo/TangZLLHLG15}.
Wai Teng Tang et al. introduce a series of bit-representation-optimized compression schemes for representing sparse matrices on GPUs including BRO-ELL, BRO-COO, BRO-HYB, which perform compression on index data and help to speed up SpMV on GPUs through reduction of memory traffic~\cite{DBLP:conf/sc/TangTRWCKGTW13}.
Jee W. Choi et al. propose a classical blocked compressed sparse row (BCSR) and blocked ELLPACK (BELLPACK) storage formats~\cite{DBLP:conf/ppopp/ChoiSV10}, which can
match or exceed state-of-the-art implementations. They also develop a performance model that can guide matrix-dependent parameter tuning which requires offline measurements and run-time estimation modelling the architecture of GPUs.
Yang et al. present a novel non-parametric and self-tunable approach~\cite{DBLP:journals/pvldb/YangPS11} to data presentation for computing this kernel, particularly targeting sparse matrices representing power-law graphs.
They take into account the skew of the non-zero distribution in matrices presenting power-law graphs.

\cparagraph{Sparse matrix storage format selection} is required because various formats have been proposed for diverse application
scenarios and computer architectures~\cite{DBLP:conf/ppopp/ZhaoLLS18}.
In~\cite{DBLP:conf/ics/SedaghatiMPPS15}, Sedaghat et al. study the inter-relation between GPU architectures, sparse matrix representation, and the sparse dataset.
Further, they build a model based on decision tree to automatically select the best representation for a given sparse matrix on a given GPU platform.
The decision-tree technique is also used in~\cite{DBLP:conf/pldi/LiTCS13}, where Li et al. develop a sparse matrix-vector multiplication auto-tuning system to bridge the gap between specific optimizations and general-purpose usage.
This framework provides users with a unified programming interface in the CSR format and automatically determines the optimal format and implementation for any input sparse matrix at runtime.
In~\cite{DBLP:conf/ppopp/ZhaoLLS18}, Zhao et al. propose to use the deep learning technique to select a right storage format.
Compared to the traditional machine learning techniques, using deep learning can avoid the difficulties in coming up with the right features of matrices for the purpose of learning.
The results have shown that the CNN-based technique can cut format selection errors by two thirds.

\cparagraph{Predictive Modeling} Machine learning has been employed for various optimization tasks~\cite{mlcpieee}, including code
optimization~\cite{wang2014integrating,Tournavitis:2009:THA:1542476.1542496,Wang:2009:MPM:1504176.1504189,wang2010partitioning,grewe2013portable,wang2013using,DBLP:journals/taco/WangGO14,taylor2017adaptive,
ogilvie2014fast,cummins2017end,ogilvie2017minimizing,ipdpsz18}, task
scheduling~\cite{grewe2011workload,emani2013smart,grewe2013opencl,ren2017optimise}, model selection~\cite{adaptivedl}, etc.

Although SpMV optimization has been extensively studied, it remains unclear how the widely-used sparse matrix representations perform on
the emerging many-core architectures. Our work aims to bridge this gap by providing an in-depth analysis on two emerging many-core
architectures (KNL and FTP), with a large number of sparse matrices and five well-recognized storage formats. Our work is the first attempt
in applying machine learning techniques to optimize SpMV on KNL and FTP.

\section{Conclusion}
This paper has presented a comprehensive study of SpMV performance on two emerging many-core architectures using five mainstream matrix
representations. We study how the NUMA binding, vectorization, and the SpMV storage representation affect the runtime performance. Our
experimental results show that the best-performing sparse matrix representation depends to the underlying architectures and the sparsity
patterns of the input datasets. To facilitate the selection of the best representation, we use machine learning to automatically learn a
model to predict the right matrix representation for a given architecture and input. Our model is first trained offline and the learned
model can be used for any \emph{unseen} input matrix. Experimental results show that our model is highly effective in selecting the matrix
representation, delivering over 90\% of the best available performance on our evaluation platforms.

\section*{Acknowledgment}
This work was partially funded by the National Natural Science Foundation of China under Grant Nos. 61602501, 11502296, 61772542, and
61561146395; the Open Research Program of China State Key Laboratory of Aerodynamics under Grant No. SKLA20160104; the UK Engineering and
Physical Sciences Research Council under grants EP/M01567X/1 (SANDeRs) and EP/M015793/1 (DIVIDEND); and the Royal Society International
Collaboration Grant (IE161012). For any correspondence, please contact Jianbin Fang (Email: j.fang@nudt.edu.cn)

\bibliographystyle{plain}
\bibliography{aspmv}

\end{document}